\documentclass[conference]{IEEEtran}
\IEEEoverridecommandlockouts
\usepackage{cite}
\usepackage{amsmath,amssymb,amsfonts}
\usepackage{algorithmic}
\usepackage{graphicx}
\usepackage{textcomp}
\usepackage{xcolor}
\usepackage{color,array,amsthm}
\usepackage{graphicx}
\usepackage{epstopdf}

\usepackage{algorithmic}
\usepackage{algorithm}
\usepackage{array}
\usepackage{svg}
\usepackage{textcomp}
\usepackage{stfloats}
\usepackage{url}
\usepackage{verbatim}
\usepackage{cite}
\usepackage{color}
\usepackage{subcaption}
\usepackage{mathtools}

\DeclarePairedDelimiter\floor{\lfloor}{\rfloor}

\def\BibTeX{{\rm B\kern-.05em{\sc i\kern-.025em b}\kern-.08em
    T\kern-.1667em\lower.7ex\hbox{E}\kern-.125emX}}
\begin{document}

\title{Deep Learning based Fast and Accurate
Beamforming for Millimeter-Wave Systems\\

\thanks{This work was supported in part by the AT\&T VURI grant. The authors would also like to thank the Army Research Labs~(ARL) and the Air Force Research Labs~(AFRL) in supporting this work through  W911NF-20-P0035 and  FA8750-20-2-0504}
}

\author{\IEEEauthorblockN{1\textsuperscript{st} Tarun S.Cousik*}
\IEEEauthorblockA{\textit{Dept. of ECE} \\
\textit{Virginia Tech}\\
Blacksburg, USA \\
tarunsc@vt.edu*}
\and
\IEEEauthorblockN{2\textsuperscript{nd} Vijay K Shah}
\IEEEauthorblockA{\textit{Department of Cybersecurity Engineering} \\
\textit{George Mason University}\\
Fairfax, USA \\
vshah22@gmu.edu}
\and
\IEEEauthorblockN{3\textsuperscript{rd} Jeffrey H Reed}
\IEEEauthorblockA{\textit{Dept. of ECE} \\
\textit{Virginia Tech}\\
Blacksburg, USA \\
reedjh@vt.edu}
\and
\IEEEauthorblockN{4\textsuperscript{th} Harry X Tran}
\IEEEauthorblockA{
\textit{ATT Labs}\\
Midtown, USA \\
xt800d@att.com}
\and
\IEEEauthorblockN{5\textsuperscript{th} Rittwik Jana}
\IEEEauthorblockA{
\textit{Google Inc}\\
NY, USA \\
rittwikj@google.com}
\and

}



\maketitle

\begin{abstract}
The widespread proliferation of mmW devices has led to a surge of interest in antenna arrays. This interest in arrays is due to their ability to steer beams in desired directions, for the purpose of increasing signal-power and/or decreasing interference levels. To enable beamforming, array coefficients are typically stored in look-up tables (LUTs) for subsequent referencing. While LUTs enable fast sweep times, their limited memory size restricts the number of beams the array can produce. Consequently, a receiver is likely to be offset from the main beam, thus decreasing received power, and resulting in sub-optimal performance. In this letter, we present BeamShaper, a deep neural network (DNN) framework, which enables fast and accurate beamsteering in any desirable 3-D direction. Unlike traditional finite-memory LUTs which support a fixed set of beams, BeamShaper utilizes a trained NN model to generate the array coefficients for arbitrary directions in \textit{real-time}. Our simulations show that BeamShaper outperforms contemporary LUT based solutions in terms of cosine-similarity and central angle in time scales that are slightly higher than LUT based solutions. Additionally, we show that our DNN based approach has the added advantage of being more resilient to the effects of quantization noise generated while using digital phase-shifters.
\end{abstract}

\begin{IEEEkeywords}
component, formatting, style, styling, insert
\end{IEEEkeywords}

\section{Introduction}

Traditionally, the idea of using large antenna arrays were limited to applications in RADAR, radio astronomy and satellite communications because of hardware complexity, application requirements and development costs\cite{ArrayHistory}. This past decade has challenged this ideology, by virtue of embracing the rapid and widespread proliferation millimeter wave(mmW) systems in commercial applications such as 5G, automative radars, fixed-satellite communication, UAV control networks, biological imaging, terrestrial mapping,in-flight entertainment services, inter-satellite communication etc \cite{imag,intro_ref}. The success of adopting the mmW regime in commercial applications as well as the lowered entry-barrier is now invigorating the development of mmW military networks for UAV swarm control, intra-vehicular LAN networks, smart warehousing, AR/VR training for mission planning and operational use, dynamic spectrum utilization, distributed command and control for Air,Space,Cyberspace etc \cite{milref,milref2}.  

Applications operating in the mmW regime typically leverage beamforming through antenna arrays for offsetting the high path loss(thus increasing the operational range), nulling interferers(in the process help with spectrum sharing and improving SINR), providing improving diversity(increasing link reliability), improving frequency reuse, and as well as the enhancing capacity gain. Beamforming, through the use of antenna arrays, has proved to be a critical enabling technology in the development of mmW applications. 

The adoption of antenna arrays as an enabling technology brings with it associated challenges and design considerations. Typically, as the number of antenna elements in the array increases, the array's ability to concentrate energy in a given direction increases. Consequently, the $3$-dB beam-width (the spatial region where the array is customarily considered effective) decreases, thus necessitating an increase in the number of beams needed to cover a given spatial region. Consequently, this leads to an increase in the cost and complexity of the hardware that creates/stores said beams, which is undesirable from a capital expenditure perspective. 

\paragraph{LUT Based Solutions}
Typically, on the fly computation of weights\footnote{In this work, weights refer to the amplitude scaling  and phase shifts applied to individual array elements} is overlooked in favor of a look-up-table (LUT) based solution to meet latency requirements \cite{Sadhu_256}. LUTs typically store a set of pre-calculated weights (also called a codebook) which correspond to a set of Beam Pointing Angles (BPAs). Codebook design (CB) is a complex application-specific task involving the determination of weights that serve requirements such as enhancing overall site coverage, minimizing drop zones, reducing beam overlap, etc \cite{CB_design}. Static-RAM (SRAM) based LUT solutions are popular analog BF solutions in conventional 5G systems where the requirements dictate the ability to scan across a small-to-moderate number of beams ($<300$) with minimal latency (reported times for pre-loading weights are $<150$ ns)\cite{CP_Leak}. However, as we continue to increase the size of our arrays and as we look at higher frequencies where the number of beams increases drastically, it is thus important to consider the limitations of LUT-based solutions. The cost of SRAM-based solutions increases in terms of implementation complexity, board area, and number of memory elements, with the number of array elements and/or the number of codewords to be loaded. \cite{CP_Leak}  \cite{ST_filter} \cite{SRAM_paper}. Consequently, this rise in costs, creates an upper bound on the size of the CB that can be loaded into the LUT, which in turn, bounds the number of serviceable BPAs for a given scan range. If asked to steer to a BPA that is unavailable in the CB, the system has two options 1) pick weights corresponding to the nearest available BPA, or 2) reload the LUT with new weights which now contain the required BPA. The first option is undesirable as it offers a sub-optimal solution to the requirement at hand. The latter is undesirable because the process of reloading the LUTs is slow thus incurring a latency penalty \cite{CP_Leak}. Moving forward, it would be advantageous to investigate approaches that circumvent the limitations of LUT based approaches while offering similar latency performance. 

\paragraph{Intelligent Analog Beamforming Solutions}
Motivated by the application of Machine Learning (ML) in tackling array challenges in direction of arrival estimation, mutual coupling reduction, EM simulation of structures etc.\cite{haupt_review}, a growing body of work has applied it to BF. For instance, in \cite{FMB_VN}, the authors evaluated the various training algorithms on two NN models. The first model was trained to predict the gain of an antenna given the weights and the second one was trained to predict the weights given the BPAs. A key finding in their work was that the Levenberg-Marquardt back propagation algorithm performed best amongst the list of training algorithms they evaluated. In the second model, the authors train a NN that estimates weights for a fixed set of BPAs but varying array sizes. In \cite{AP2In_NN}, the authors built a network that predicts the weights for a 4-element linear array, after it is fed with the radiation pattern. To demonstrate the efficacy of their solution, they overlayed an instantiation of a normalized pattern generated from the weights in the validation set, and showed that it fitted well over the expected pattern. Authors in \cite{CNN_A2P2In} investigated the use of CNNs in predicting BF coefficients. The CNN had two input channels, corresponding to images of the pattern in the linear and dB domain. These inputs were passed through the convolutional layers and then through a FNN in order to regress to the corresponding weights. The authors then visually compared the 2-D patterns generated by the CNN's predicted weights versus the actual pattern and concluded that an acceptable performance was achieved. In \cite{NetArch_Comp}, the authors investigated how the choice of network architecture affects accuracy and latency in providing BF solutions. In their work, they found that the FNN-based architectures offered the lowest latency ($2-5\times$) when compared to CNNs, Gated Recurrent Unit (GRU), and Long Term Short Term Memory(LSTM) networks. Furthermore, in the performance phase, when comparing the mean squared error~(MSE) of the real and imaginary components of the predicted weights, they observed that the MSE of the FNN solution was only ~$6\%$ worse than the best performing GRU-NN. After considering the latency and performance comparisons in \cite{NetArch_Comp}, we picked the FNN as the candidate architecture for our work.

\paragraph{Limitations of Previous Work and Our Contributions}
Existing intelligent beamforming solutions serve two purposes: 1) They motivate the use of NN in BF applications 2) Provide some key insights in building NN-based BF systems. These existing work would further benefit in making their case from highlighting the following points:
\begin{itemize}
\item In \cite{FMB_VN,AP2In_NN,CNN_A2P2In}, the authors focus their contribution to a narrow range of scanning angles.This would be undesirable in practical situations where a wide operational range is required. This issue is rectified in our work by showcasing the performance of our solution in a typical 3-D sector.
\item In~\cite{FMB_VN,AP2In_NN,NetArch_Comp}, the work is applied to small linear arrays($<8$ elements). Limiting analysis to small arrays, brings into question the applicability of ML-based BF when the array size is scaled to larger planar arrays. Additionally, a consequence of developing solutions to linear arrays limits their scanning range to the 2-D plane. In this work, we develop our solution for a 2-D array(with 64 array elements) thus, verifying the scalability of ML-based BF solutions and also enabling 3-D scanning.
\item In \cite{FMB_VN,AP2In_NN,CNN_A2P2In}, the performance of the ML based BF solutions are demonstrated by overlaying antenna patterns on top of each other, thus rendering the interpretation of performance to be very reader specific/subjective. Additionally, since only a few patterns can be visually overlayed on top of each other due to space constraints, it  prevents ~\cite{FMB_VN,AP2In_NN,CNN_A2P2In} from showcasing the robustness of their proposed approaches or from demonstrating how performance divergences across various BPAs. In this work, we leverage performance metrics such as BPA deviation (measured in terms of Central Angle) and Beam pattern similarity (measured in terms of cosine similarity) to show the effectiveness of our proposed solution over an extensive range of beam pointing angles.

\item In~\cite{FMB_VN,AP2In_NN,CNN_A2P2In,NetArch_Comp}, since the authors do not compare their solution's performance against the conventional CB-based approaches, there is no demonstrable motivation provided to array designers to switch away from a CB based approach to a DNN based solution. This is a \textbf{crucial detriment} to an otherwise excellent idea. Thus in thus work, we remedy this issue by drawing a direct comparison of the effectiveness of our approach against conventional CB-based designs. By doing so, we ascertain certain trade-offs that were otherwise not observed in previous work. 

\item In\cite{FMB_VN,AP2In_NN,CNN_A2P2In,NetArch_Comp}, the authors have limited their solutions to ideal analog phase shifters; since real-world beamformers are often implemented with digital phase shifters, the impact of phase-quantization on DNN based solutions is not well characterized. In our work, we showcase how our solution performs with the addition of  quantization error introduced by the use of digital phase shifters. We additionally show our DNN-based solution performs significantly better than CB based approach with similar constraints. In fact, we show for the first time, that DNN-based solutions are relatively immune to quantization noise compared to a typical CB-based approach and possible reasons for this are discussed.

\end{itemize}

The outline of this work is as follows: Section II discusses the system model and baseline codebook approach. Section III present the BeamShaper framework. In section IV, we discuss the evaluation metrics followed by performance analysis in Section V. Section VI concludes the paper.


\section{System Model}
We consider a mmW network that consists of a directional transmitter and multiple receivers that are scattered randomly around the transmitter. We assume the transmitter has a planar array composed of $N$ electrically small-dipole antenna elements, spaced half wavelength apart. The transmitter's objective is to steer the main beam to a given user. The parameters $\phi$ and $\theta$ are used to represent a particular instance of azimuth (Az) and elevation (El) angle the main beam of the array points to. $\phi$ and $\theta$ together represent the BPA. We assume a total of $m$ BPAs to train our network and evaluate its performance over. The  weights for each element for the baseline are generated using the maximum gain beamforming (MGB)/phase-only beamforming criteria and can be given by the following expression \cite{balanis}: 
\begin{equation}
I_{{n}_{MGB}}(\phi,\theta) = e^{-j\beta P_{n}\cdot(  \hat{x} \sin(\theta) \cos(\phi) + \hat{y} \sin(\theta) \sin(\phi) + \hat{z} \cos(\theta) )}
\label{EqnMGB}
\end{equation}
where $I_{n}$ represents the weight loaded to the $n^{th}$ array element for a BPA given by $(\theta,\phi)$, $n \in [0,N-1]$ and  $P_{n}$ represents the $n^{th}$ element's position relative to the $0^{th}$ element and $\beta$ is the wave number. 

We picked the IEEE $802.11.15.3$ codebook as a baseline to evaluate the performance of BeamShaper due to its popularity as a baseline and its effectiveness to work relatively well when the number of beams to cover a region is relatively large \cite{CB_design}. This CB allows the designer to set and/or vary the size of the CB (K). Let $k \in [0,K-1]$ represent an arbitrary beam within the CB. The weights loaded to the $n^{th}$ element for the $k$-{th} beam is given by:
\begin{equation}
I_{{n}_{CB}}(k) = e^{j \frac{2\pi}{2^b} \rceil\frac{n \times mod[k-1 + (K/2),K]}{\frac{K}{2^{b}}}\rceil}
\label{EqnCBK}
\end{equation}
where $b$ is used to represent the resolution of the phase shifter. We would also like to note, while it is a somewhat easy code-book to compute, it does not offer the designer any direct control over where the designed beams point.
\vspace{-0.15 cm}

\section{BeamShaper Framework}
As shown in Figure \ref{fig:sys}, the proposed BeamShaper framework employs a pretrained deep learning model that inputs a desired beam pointing angle (BPA) as tuple $<\theta, \phi>$ (where $\theta$ and $\phi$ respectively are azimuth and elevation angle.); and outputs on-the-fly the exact antenna weights for each array element, such that, the base station creates a main beam lobe in the inputted desired BPA direction. The BeamShaper framework is envisioned to be stationed on a computational node, such as, a FPGA, GPU, TPU or ASIC located at the base station.
\begin{figure}
    \centering
    \captionsetup{justification=centering}
	\centerline{\includegraphics[ width=.5\textwidth]{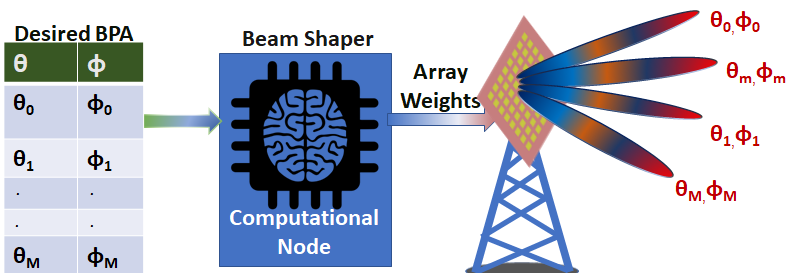}}
	\caption{Overview of BeamShaper Framework}
	\label{fig:sys}
\end{figure}

\paragraph{Architecture Specifics}
The DL architecture used for BeamShaper is a Feed-Forward Network with three hidden layers comprising of $[32,300,600]$ neurons, respectively. The selection of this architecture was motivated from previous works\cite{NetArch_Comp} that showed that a FNN architecture provided the best performance when considering latency constraints into account. The number of neurons in each layer was determined by attempting various combinations of neurons in each layer and picking the architecture with the lowest validation loss. The network is fed with two inputs that correspond to normalized beam pointing angles in $\phi$ and $\theta$ (discussed in more detail in \ref{Data_Processing}). The output of the network is $64$ neurons and each neuron provides the phase applied to individual elements of the array pointing in the direction of $\phi$ and $\theta$. The challenge of finding appropriate weights is thus reformulated as a Neural Network based regression problem. 

The first two layers are provided with the Snake and TSigmoid activation functions~\cite{snek}. The batch size is set to $1024$ and the network is trained with the ADAM optimizer \cite{adam}. The loss function used for training was MSE. A variety of learning rates were tested and we found $0.0005$ to be optimal when used with $1200$ epochs. Table I, summarizes the various  parameters that are used to train and develop BeamShaper.

\paragraph{Dataset Preprocessing and Training Procedures}\label{Data_Processing}
The weights calculated using the MGB solution in Eq.~\ref{EqnMGB} have an absolute value of $1$. We take advantage of this fact and convert the complex weights to their argument (in degrees). We found that feeding the input weights in normalized polar form slightly increased BeamShaper's accuracy (not explicitly shown). The data-set comprising of the BPAs and the corresponding  weights  are normalized using SKLearn's Standard Scaler \cite{scikit} to values between $[-1,1]$. The dataset is then partitioned three ways into training, validation and testing with the ratio of $70/15/15$. 

\paragraph{Evaluation Metrics}
Since, our end goal is to evaluate the efficacy of BeamShaper in terms of radiation performance, this subsection develops two key performance metrics for this purpose, namely, BPA Deviation and Beam Pattern Similarity, that can be used to evaluate the effectiveness of our BF solution. 


\underline{\textit{BPA Deviation:}} When the beam produced by an array system is offset from the desired direction, it reduces the received power and thus causes a reduction in receiver performance. It is thus desirable to minimize the deviation in the BPA (henceforth referred to as \textit{BPA deviation}). To quantify the  BPA deviation BPA  in either the BeamShaper or the CB relative to the MGB solution, we leverage the use of the \textit{Central Angle}(CA) which can be obtained from the great circle distance. The great circle distance (shown in magenta in Fig 2) measures the shortest distance between two points on the surface of a sphere, measured along the surface of the sphere. The Central Angle measures the angle enclosed between the great circle distance and the origin (shown with a green dot). The great circle distance and the Central Angle are popular measures used in the geospatial community to map distances between various locations under a spherical earth assumption. The Central Angle(denoted by $\alpha$) for two arbitrary points $\textbf{A}$ and $\textbf{B}$ located on the surface of a sphere at locations $(\phi_A,\theta_A)$ and $(\phi_B,\theta_B)$ can be given by the following formula\footnote{This formula is susceptible to large rounding errors if a low floating-point precision is used. We found that using a 64 bit floating-point showed no signs of rounding errors in our work.}~\cite{Cent_Ang}:
\begin{equation}
    \alpha = \cos{\theta_{A}}\cos{\theta_B} +\sin{\theta_A}\sin{\theta_B}\cos(\phi_A -\phi_B) 
\end{equation}
\begin{figure}
    \centering
	\centerline{
 \includegraphics[ width=.35 \textwidth]{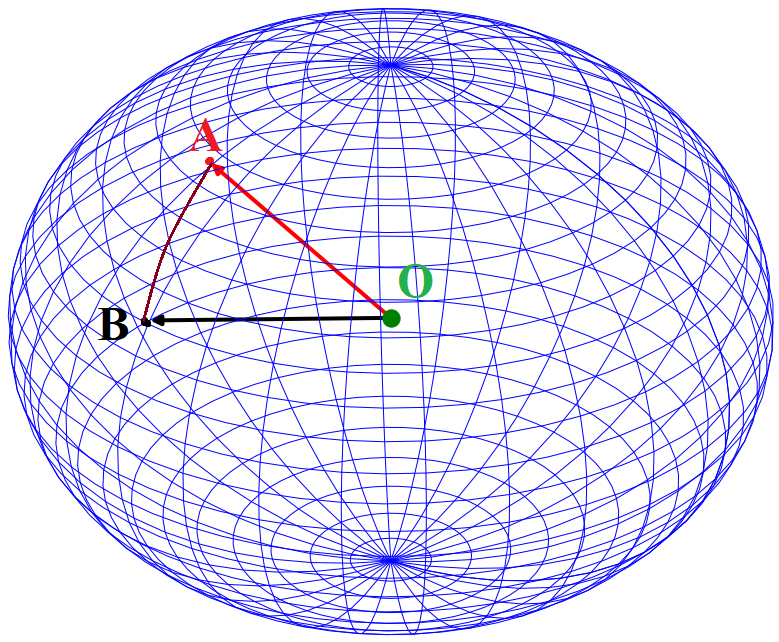}}
	\caption{Central Angle and Great Circle Distance }
	\label{fig:sys}
\end{figure}
\vspace{-0.6cm}

Ideally, the CA should be $0$ and is attained when the array's main beam points exactly in the direction that is required. Thus, if all the BPAs of a given approach perfectly line up with that of the MGB approach, the CDF would resemble an unit impulse function centred at $CA = 0$ with a height of $1$. 

\underline{\textit{Beam pattern Similarity:}}
By re-imagining the radiation patterns as a $D$-dimensional vector (where $D$ is the number of sample points used to represent a pattern), we can use Cosine Similarity ($CSim$) to compute how ``close" two given patterns are relative to one another in an inner-product sense\cite{CSimRef}, referred to as, \textit{Beam pattern Similarity}. This is because $CSim$ between two patterns/vectors is their normalized inner product and can be calculated as follows:
\begin{equation}
CSim_{NN/CB} = \frac{ F_{MGB} \cdot  F_{NN/CB}}{||F_{MGB}||_{2}  \,\, ||F_{NN/CB}||_{2}}
\end{equation}
where, $F_{MGB}$ , $F_{NN}$, $F_{CB}$ are the radiation patterns from the MGB, BeamShaper and CB approaches, respectively, and $||X||_{2}$ represents the $\ell^2$-norm of any vector $X$. 

Properties that make $CSim$ a desirable metric include reflexivity, symmetricty, and shift-invariance. $CSim$'s shift variance property ensures that we capture instances where patterns that are otherwise identical are shifted spatially. Patterns are considered identical if their $CSim$score is $1$ and are  dissimilar if the $CSim$ score is 0 .
\vspace{-0.3cm}

\section{Performance Analysis}
\paragraph{Simulation Settings}
We assumed that the transmitter operates over a spatial coverage region of $\phi \in U[0,120]$ and $\theta \in U[30,150]$, where $U[.]$ refers to a uniform distribution. This coverage region corresponds to a typical cellular sector that directional eNodeBs operate on. We generated $m = 1.5\times 10^6$ BPAs within the coverage regions for the purpose of training and testing BeamShaper against a CB-based approach. The transmit array at the BeamShaper is assumed to have $N=64$ elements. We calculate the MGB solution for all $m$ BPAs using \ref{EqnMGB}. The IEEE $802.11.15.3$ CB is implemented for the same array using \ref{EqnCBK}, and its size, K, is picked from a set given by $K \in [16, 64, 256,1024] $. While contemporary LUT-based architectures at the time of this work  are typically restricted to CB sizes of $256$  \cite{CP_Leak}\cite{Sadhu_256}, we included $1024$ in our analysis to highlight the boost in performance offered by BeamShaper. In our initial simulations, we initially disregard the phase shifter's resolution and set the value of $b$ to an arbitrarily large value ($>15$). We then find the codebook that is closest to the BeamShaper's performance and compare their performance on both $CSim$ and $CA$ for $b\in {1,2,..6}$.Table II summarizes various simulation parameters used in this work.

\begin{table*}
\parbox{.4\linewidth}{
\caption{Simulation Parameters}
\centering
\begin{tabular}{|l|l|}
        \hline
        
        \textbf{Parameter} & \textbf{Value}\\
        \hline
        Array Size & $64$\\
        \hline
        Az Distribution & U[$0$,$120$] \\
        \hline
        El Distribution & U[$30$,$150$] \\
        \hline
        Number of BPAs & $1.5 \times 10^6$ \\
        \hline
        Number of codewords & $16,64,256,1024$ \\
        \hline
        Train/Validation/Test Split & $70/15/15$\\
        \hline
        
    \end{tabular}
}
\hfill
\parbox{.6\linewidth}{
\centering
\caption{Quartile Comparison of BeamShaper against the Codebook}
\begin{tabular}{|l|l|l|l|l|l|l|l|l|}
   \hline
\textbf{Approach}  & \multicolumn{4}{c|}{\textbf{BPA Deviation}\textbf{(deg$^{\circ}$)}}
            & \multicolumn{4}{c|}{\textbf{Pattern Integrity}}
            \\

              & \textbf{$p25$}
              & \textbf{$p50$}  
              & \textbf{$p75$} 
              & \textbf{$p95$} 
              & \textbf{$p25$}  
              & \textbf{$p50$}  
              & \textbf{$p75$}  
              & \textbf{$p95$}  \\
    \hline
Codebook $16$   & $13.2$&  $19.2$ & $23.3$ & $30.6$  &   $0.582$ &  $0.774$   &  $0.918$   & $0.986$    \\
    \hline
Codebook $64$   & $5.7$ &  $8.3$     &  $12.0$     &  $20.1$     & $0.849$      & $0.933$    & $0.975$      &  $0.995$      \\
    \hline
Codebook $256$   & $2.9$    & $4.2$    &  $6.2$   & $11.3$    & $0.945$    & $0.973$    &  $0.989$   & $0.998$    \\
    \hline
Codebook $1024$   & $1.4$      & $2.5$      & $4.0$    &  $8.3$     &  $0.979$     &  $0.990$     & $0.996$      &   $0.999$    \\
    \hline

\textbf{BeamShaper} & \textbf{$0.7$} & \textbf{$1.3$}& \textbf{$2.7$}&\textbf{$6.8$}& \textbf{$0.994$}&\textbf{$0.996$}& \textbf{$0.998$}&\textbf{$0.999$}\\
    \hline

    \end{tabular} 
}
    \end{table*}


\paragraph{Results}
\begin{figure*}[ht]
    \centering
    \includegraphics[width=\textwidth]{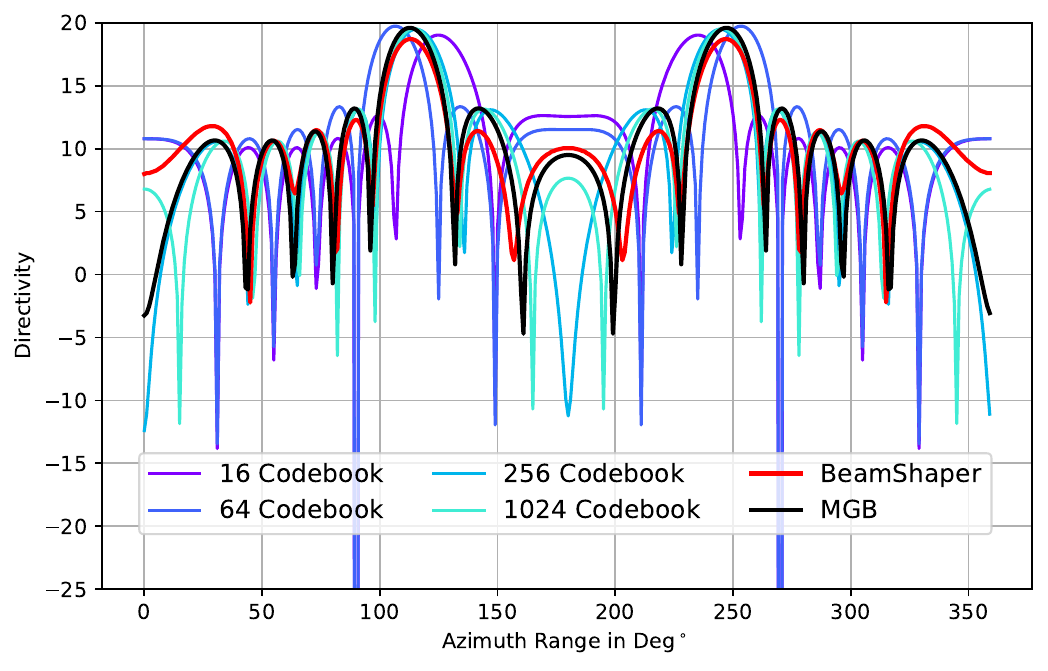}
   \caption{Comparison of Directivity generated through various approaches for a random BPA}
   \label{fig:Pattern_Comp}
\end{figure*}

\begin{figure*}[ht]
    \centering
    \includegraphics[width=\textwidth]{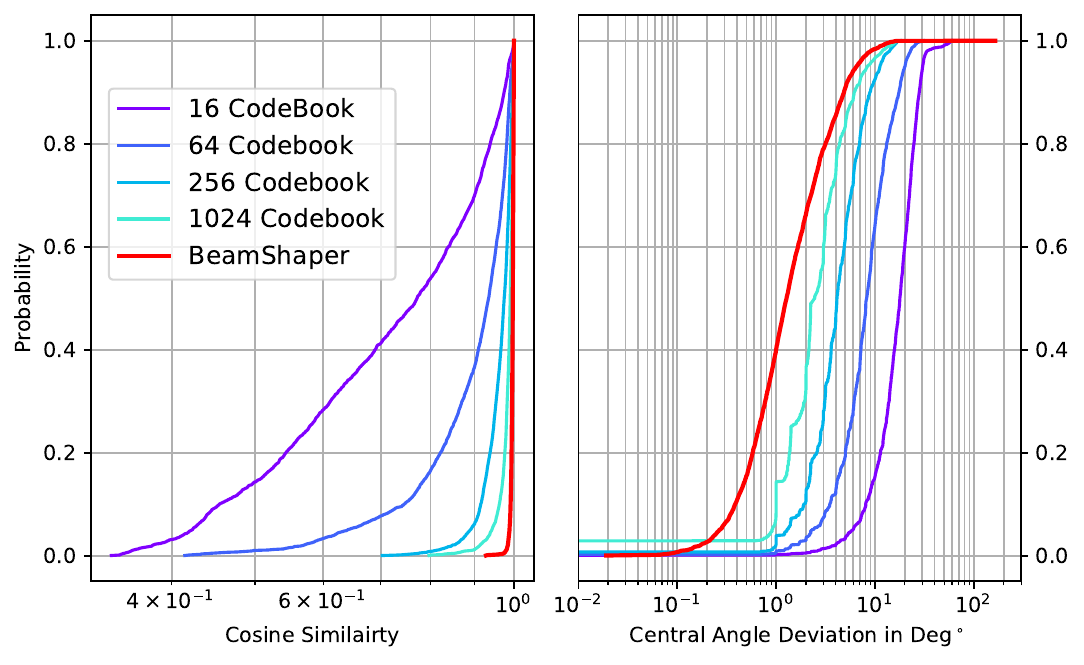}
   \caption{ Comparison of CDF of Cosine Similarity, and BPA Deviation}
   \label{fig:CS}
   
\end{figure*}
A visual inspection of  Fig \ref{fig:Pattern_Comp} reveals that the weights produced by BeamShaper are able to generate a pattern that closely resembles to the radiation pattern obtained from the MGB solution. Once trained, BeamShaper approach is able to produce the weights for a given azimuth and elevation angle on an average of $270$ ns.

Results from Fig \ref{fig:Pattern_Comp} and Fig \ref{fig:CS} show that BeamShaper outperforms even the best CB based strategy in all quartiles of both CA and $CSim$. In fact, when observing the various quartiles, we observe that BeamShaper on average produces a CA that is roughly half of what is produced by the $1024$ sized CB. Likewise, we note that every quartile of the BeamShaper consistently performs as well or outperforms  the next quartile of the $1024$ beam CB. Table 3 summarizes BeamShaper's performance against the various CB sizes.

We now pick the best performing codebook ($1024$) and compare it against BeamShaper after introducing phase quantisation. Phase Quantisation is introduced into real-world arrays by using digital phase shifters (DPS) in the BF process. DPS have a fixed number of phase-states they can toggle between and this number is limited by the bit resolution of the DPS. In our analysis, we vary $b$ between $2$ and $6$ bits. This range emulates typical DPS bit-resolution encountered in the real world. The weights of the individual antenna weights after quantization can be determined in the following manner:
{\begin{equation}
I_{{nq}_{CB/NN}} = \floor*{ \frac{I_{{n}_{CB/NN}}}{\Phi_{R}} + \frac{1}{2}}\times{\Phi_{R}}
\end{equation}}
{where $\Phi_{R}$ is the phase resolution given by $\frac{360}{2^b}$}

\begin{figure*}[ht]
    \centering
    \includegraphics[width=\textwidth]{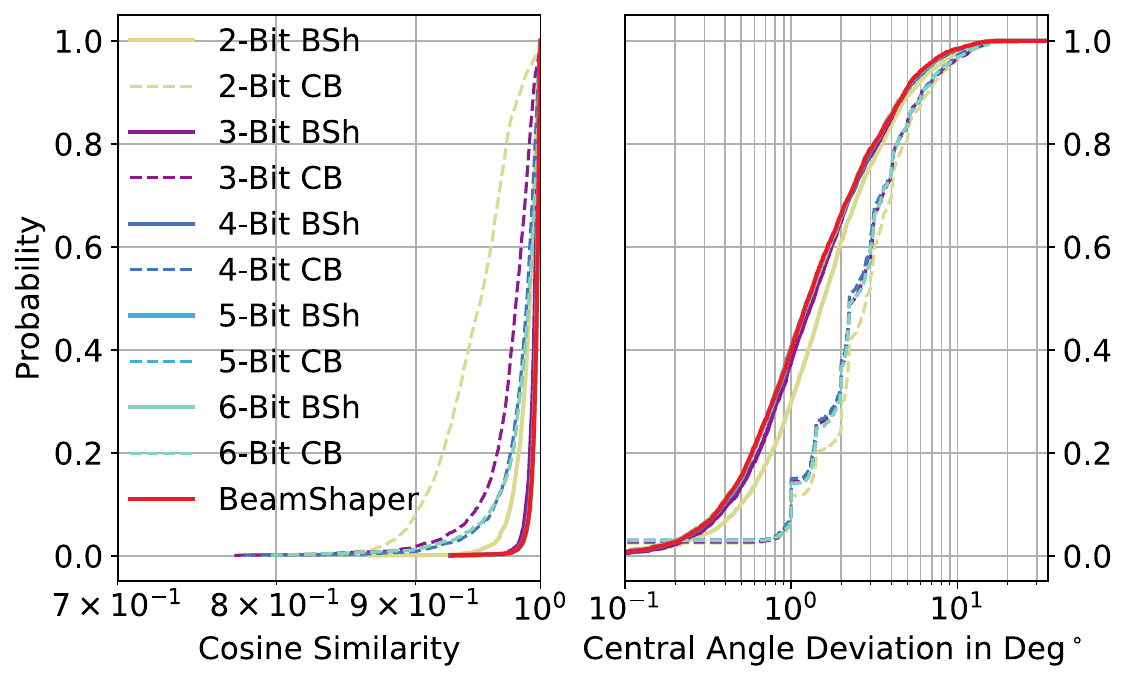}
   \caption{ Effect of Phase Quantization on CDF of Cosine Similarity and BPA Deviation for 1024-CB and BeamShaper(BSh)}
\end{figure*}

In Fig 5, we note that the deleterious effects of quantization is more pronounced in the case of the codebook. 
This difference in performance can be attributed to the product found by combining the phase quantization error(common to both approaches) and the  misalignment error in the weights due to the  initial beam misalignment. This misalignment error is relatively excessive in the CB based approach, due to the limited memory. Consequently, this results in the CB's performance dropping faster when the bit resolution is decreased.  It is worth noting that a 2-bit phase shifter can adversely  affect even the extremely large 1024-CB in terms of CSim and CAD. It is also worth pointing out that BeamShaper with a 2-bit phase resolution manages to perform as well as the 1024-codebook with a 6-bit resolution in terms of CSim and significantly outperforms the 6 bit 1024-CB in terms of CA. Consequently, we infer, that in addition to enhancing performance with regards to CSim and CA, BeamShaper gives the array designer the added benefit of leveraging cheaper phase shifters of lower resolution while delivering equal if not greater performance relative to the CB based design. 

\paragraph{Limitations and Discussion}
In our present analysis, we have not accounted for mutual coupling; the reason for this is two-fold:
1. Array designers routinely add structures that mitigate/eliminate mutual-coupling through structures like inset lines, defected ground planes, parasitic elements  etc \cite{MC_red}. We thus assume, that such mitigation techniques could be used in a a more realistic array that uses BeamShaper.
2. Our aim was to document the difference phase-quantization introduced between a DNN-based BF approach and a CB-based approach in a isolate fashion. The introduction of mutual coupling would complicate this as it would then have to separately account for coupling based effects vs those introduced by phase-quantization. Future studies may include mutual coupling to understand the 
impact on performance. 
While we observed the performance of BeamShaper to outperform LUT based approach in terms of $CSim$ and CA in an MGB setting, the applicability of DNN-based approaches to produce multi-beamforming solutions is yet to be studied in a rigorous manner.  Our current architecture calculates the weights in about $270$ ns, which is slightly more than the switching speeds seen in LUT-based approaches(4-150 ns)~\cite{CP_Leak}. If this additional delay(~100-200ns) can be tolerated, BeamShaper provides an  attractive alternative that greatly boosts radiation performance. Further research is required to understand if the computational latency can be further reduced by including dropout layers as well as by exploring alternative architectures.

Our current solution was implemented on a dedicated GPU(16 GB Nvidia v100). More studies are needed to understand how BeamShaper's translates implemented on FPGAs, ASICs or other embedded platforms. Lastly, since our current work does not take into account real-world effects quantization noise from ADC/DACs or phase noise generated at the Tx/Rx additional insights could be obtained through the inclusion of these effects.

\section{Conclusion}
In this work, we proposed and investigated the use of BeamShaper, a Deep Neural Network designed for generating beamforming weights in mmW antenna array systems. We compared the performance of BeamShaper against popular codebook-based approaches of various sizes. Our results showed that BeamShaper outperforms even the best codebook-based baseline in terms of central angle and cosine similarity with both ideal and realistic phase shifters. We found that this performance boost came at the cost of slighlt increased computationally latency. We thus concluded that in applications where performance dictates requirements, BeamShaper and other NN based BF solutions offer some attractive benefits compared to LUT based solutions. 
\vspace{-0.3cm}

\bibliographystyle{unsrt}
\bibliography{beamshaper.bib}

\begin{thebibliography}{10}

\bibitem{ArrayHistory}
Randy~L. Haupt and Yahya Rahmat-Samii.
\newblock Antenna array developments: A perspective on the past, present and
  future.
\newblock {\em IEEE Antennas and Propagation Magazine}, 57(1):86--96, 2015.

\bibitem{imag}
Roger Appleby and Rupert~N. Anderton.
\newblock Millimeter-wave and submillimeter-wave imaging for security and
  surveillance.
\newblock {\em Proceedings of the IEEE}, 95(8):1683--1690, 2007.

\bibitem{intro_ref}
Table of frequency allocations chart, Jul 2016.

\bibitem{milref}
James~F. Harvey, Michael~B. Steer, and Theodore~S. Rappaport.
\newblock Exploiting high millimeter wave bands for military communications,
  applications, and design.
\newblock {\em IEEE Access}, 7:52350--52359, 2019.

\bibitem{milref2}
Dod announces \$600 million for 5g experimentation and testing at five
  installations.

\bibitem{Sadhu_256}
Bodhisatwa Sadhu, Yahya Tousi, Joakim Hallin, Stefan Sahl, Scott Reynolds,
  {\"O}rjan Renstr{\"o}m, Kristoffer Sj{\"o}gren, Olov Haapalahti, Nadav Mazor,
  Bo~Bokinge, et~al.
\newblock 7.2 a 28ghz 32-element phased-array transceiver ic with concurrent
  dual polarized beams and 1.4 degree beam-steering resolution for 5g
  communication.
\newblock In {\em 2017 IEEE International Solid-State Circuits Conference
  (ISSCC)}, pages 128--129. IEEE, 2017.

\bibitem{CB_design}
Jianhua Mo, Boon~Loong Ng, Sanghyun Chang, Pengda Huang, Mandar~N. Kulkarni,
  Ahmad Alammouri, Jianzhong~Charlie Zhang, Jeongheum Lee, and Won-Joon Choi.
\newblock Beam codebook design for 5g mmwave terminals.
\newblock {\em IEEE Access}, 7:98387--98404, 2019.

\bibitem{CP_Leak}
Jian Pang, Zheng Li, Xueting Luo, Joshua Alvin, Kiyoshi Yanagisawa, Yi~Zhang,
  Zixin Chen, Zhongliang Huang, Xiaofan Gu, Weichu Chen, Yun Wang, Dongwon You,
  Zheng Sun, Yuncheng Zhang, Hongye Huang, Naoki Oshima, Keiichi Motoi,
  Shinichi Hori, Kazuaki Kunihiro, Tomoya Kaneko, Atsushi Shirane, and Kenichi
  Okada.
\newblock A fast-beam-switching 28-ghz phased-array transceiver supporting
  cross-polarization leakage self-cancellation.
\newblock In {\em 2021 Symposium on VLSI Circuits}, pages 1--2, 2021.

\bibitem{ST_filter}
Arun Paidimarri and Bodhisatwa Sadhu.
\newblock Spatio-temporal filtering: Precise beam control using fast beam
  switching.
\newblock In {\em 2020 IEEE RFIC Symposium}, pages 207--210, 2020.

\bibitem{SRAM_paper}
Erik Öjefors, Mikael Andreasson, Torgil Kjellberg, Håkan Berg, Lars Aspemyr,
  Richard Nilsson, Klas Brink, Robin Dahlbäck, Dapeng Wu, Kristoffer Sjögren,
  and Mats Carlsson.
\newblock A 57-71 ghz beamforming sige transceiver for 802.11 ad-based fixed
  wireless access.
\newblock In {\em 2018 IEEE (RFIC Symposium}, pages 276--279, 2018.

\bibitem{haupt_review}
Francesco Zardi, Payam Nayeri, Paolo Rocca, and Randy Haupt.
\newblock Artificial intelligence for adaptive and reconfigurable antenna
  arrays: A review.
\newblock {\em IEEE Antennas and Propagation Magazine}, 63(3):28--38, 2021.

\bibitem{FMB_VN}
Theodoros~N Kapetanakis, Ioannis~O Vardiambasis, George~S Liodakis, Melina~P
  Ioannidou, and Andreas~M Maras.
\newblock Smart antenna design using neural networks.
\newblock In {\em 8th International Conference: New Horizons in Industry,
  Business and Education (NHIBE 2013)}, pages 130--135, 2013.

\bibitem{AP2In_NN}
Jae~Hee Kim and Sang~Won Choi.
\newblock A deep learning-based approach for radiation pattern synthesis of an
  array antenna.
\newblock {\em IEEE Access}, 8:226059--226063, 2020.

\bibitem{CNN_A2P2In}
Ricardo Lovato and Xun Gong.
\newblock Phased antenna array beamforming using convolutional neural networks.
\newblock In {\em 2019 IEEE International Symposium on Antennas and Propagation
  and USNC-URSI Radio Science Meeting}, pages 1247--1248, 2019.

\bibitem{NetArch_Comp}
Haya Al~Kassir, Zaharias~D. Zaharis, Pavlos~I. Lazaridis, Nikolaos~V.
  Kantartzis, Traianos~V. Yioultsis, Ioannis~P. Chochliouros, Albena Mihovska,
  and Thomas~D. Xenos.
\newblock Antenna array beamforming based on deep learning neural network
  architectures.
\newblock In {\em 2022 3rd URSI Atlantic and Asia Pacific Radio Science Meeting
  (AT-AP-RASC)}, pages 1--4, 2022.

\bibitem{balanis}
Constantine~A Balanis.
\newblock {\em Antenna theory:Analysis and Design, John Wiley \& sons}.
\newblock 2015.

\bibitem{snek}
Liu Ziyin, Tilman Hartwig, and Masahito Ueda.
\newblock Neural networks fail to learn periodic functions and how to fix it.
\newblock {\em Advances in Neural Information Processing Systems},
  33:1583--1594, 2020.

\bibitem{adam}
Diederik~P Kingma and Jimmy Ba.
\newblock Adam: A method for stochastic optimization.
\newblock {\em arXiv preprint arXiv:1412.6980}, 2014.

\bibitem{scikit}
F.~Pedregosa, G.~Varoquaux, A.~Gramfort, V.~Michel, B.~Thirion, O.~Grisel,
  M.~Blondel, P.~Prettenhofer, R.~Weiss, V.~Dubourg, J.~Vanderplas, A.~Passos,
  D.~Cournapeau, M.~Brucher, M.~Perrot, and E.~Duchesnay.
\newblock Scikit-learn: Machine learning in python.
\newblock {\em Journal of Machine Learning Research}, 12:2825–2830, 2011.

\bibitem{Cent_Ang}
Lyman~M Kells.
\newblock {\em Plane and Spherical Trigonometry with Tables by Lyman M. Kells,
  Willis F. Kern, James R. Bland}.
\newblock McGraw Hill Book Company, 1940.

\bibitem{CSimRef}
Pang-Ning Tan, Michael Steinbach, and Vipin Kumar.
\newblock {\em Introduction to data mining}.
\newblock Pearson Education India, 2016.

\bibitem{MC_red}
Iram Nadeem and Dong-You Choi.
\newblock Study on mutual coupling reduction technique for mimo antennas.
\newblock {\em IEEE Access}, 7:563--586, 2019.

\end{thebibliography}
\end{document}